\newcommand{\ifb}{\ensuremath{\mathrm{fb^{-1}}}}
\newcommand{\TeV}{\ensuremath{\mathrm{Te\kern -0.1em V}}}
\newcommand{\TeVc}{\ensuremath{\mathrm{Te\kern -0.1em V\!/}c}}
\newcommand{\TeVcc}{\ensuremath{\mathrm{Te\kern -0.1em V\!/}c^2}}
\newcommand{\GeV}{\ensuremath{\mathrm{Ge\kern -0.1em V}}}
\newcommand{\GeVc}{\ensuremath{\mathrm{Ge\kern -0.1em V\!/}c}}
\newcommand{\GeVcc}{\ensuremath{\mathrm{Ge\kern -0.1em V\!/}c^2}}
\newcommand{\MeV}{\ensuremath{\mathrm{Me\kern -0.1em V}}}
\newcommand{\MeVc}{\ensuremath{\mathrm{Me\kern -0.1em V\!/}c}}
\newcommand{\MeVcc}{\ensuremath{\mathrm{Me\kern -0.1em V\!/}c^2}}

\newcommand{\babar}{\mbox{\slshape B\kern-0.1em{\small A}\kern-0.1em B\kern-0.1em{\small A\kern-0.2em R}}}

\def\massdifffitone{1046.7^{+2.9}_{-3.0}~\MeVcc }
\def\widthfitone{15.3^{+10.4}_{-6.1}~\MeVcc }

\def\massone{4143.4^{+2.9}_{-3.0}(\mathrm{stat})\pm0.6(\mathrm{syst})~\MeVcc }
\def\widthone{15.3^{+10.4}_{-6.1}(\mathrm{stat})\pm2.5(\mathrm{syst})~\MeVcc }
\def\yieldone{19\pm6 }
\def\rebfone{0.149\pm0.039(\mathrm{stat})\pm0.024(\mathrm{syst}) }

\def\masstwo{4274.4^{+8.4}_{-6.7}(\mathrm{stat})~\MeVcc }
\def\widthtwo{32.3^{+21.9}_{-15.3}(\mathrm{stat})~\MeVcc }
\def\yieldtwo{22\pm8 }

\documentclass{PoS}

\title{Observation of a Narrow Near-Threshold Structure in the $J/\psi\phi$ Mass Spectrum in $B^+\to J/\psi\phi K^+$ Decays}

\ShortTitle{35th ICHEP}

\author{\speaker{Kai Yi}\thanks{for the CDF Collaboration}\\
        University of Iowa\\
        E-mail: \email{yik@fnal.gov}}

\abstract{
Observation is reported for a structure  near  the  $J/\psi\phi$ threshold  
in  $B^+\rightarrow J/\psi\phi K^+$ decays produced in $\bar{p} p $
collisions at $\sqrt{s}=1.96~\TeV$  with a statistical significance of   
beyond 5 standard deviations. There are $\yieldone$ events observed 
for this structure at a mass of $\massone$ and a width of $\widthone$, 
which are consistent with the previous measurements reported as 
evidence of the $Y(4140)$.           

}

\FullConference{35th International Conference of High Energy Physics - ICHEP2010,\\
		July 22-28, 2010\\
		Paris France}

\begin{document}

Recently, evidence  has been reported by CDF for a narrow  structure 
near the  $J/\psi\phi$ threshold, named  $Y(4140)$,  
in  $B^+\rightarrow J/\psi\phi K^+$ decays produced in $\bar{p} p $
collisions at $\sqrt{s}=1.96~\TeV$~\cite{y4140}.  
The structure is the first charmonium-like structure decaying into two heavy 
quarkonium states ($c\bar{c}$ and $s\bar{s}$) which is a candidate for 
exotic mesons~\cite{y4140theory}. 
In this note, we report an update on a search for structures in 
the $J/\psi\phi$ system 
produced in exclusive $B^+\rightarrow J/\psi\phi K^+$ decays 
with $J/\psi \rightarrow \mu^+ \mu^-$ and $\phi\rightarrow K^+K^-$. 
This analysis is based on a sample  of $\bar{p} p $
collision data at $\sqrt{s}=1.96~\TeV$ with an integrated luminosity of about 
6.0 $\ifb$ collected by the CDF II detector at the Tevatron. 
The CDF II detector has been described in detail elsewhere~\cite{cdfii}. 
In this analysis, $J/\psi\to\mu^+\mu^-$ events are recorded 
using a dedicated three-level dimuon trigger.

The invariant mass of $J/\psi\phi K^+$ in the current dataset, which includes 
those used in the previous analysis 
after applying the same requirements used in 
the previous analysis~\cite{y4140}, is shown in Fig.~\ref{myfig1}. 
A fit with a Gaussian signal function with its rms fixed to  
the value 5.9 \MeVcc~obtained from Monte Carlo (MC) simulation~\cite{cdfbc} 
and a linear background function  to the mass spectrum of $J/\psi\phi K^+$  
returns a $B^+$ signal of $115\pm12(\mathrm{stat})$ events.
For a luminosity increase by a factor of 2.2, the yield increase was 1.53,
reduced by trigger rate-limitation at higher average luminosity.
We then select   $B^+$ signal candidates with a mass 
within  3$\sigma$ (17.7 \MeVcc) of the nominal $B^+$ mass. 
We define those events with a mass within [-9,-6]$\sigma$ or [6,9]$\sigma$ 
of nominal $B$ mass as B sideband events. 
Fig.~\ref{myfig2} shows 
the mass difference, $\Delta M= m(\mu^+\mu^-K^+K^-)- m(\mu^+\mu^-)$, for events in 
the $B^+$ mass window as well as in the B sideband in our data sample. 
The comparison of the $\Delta M$ distribution in the B mass window for the dataset used in this
analysis (6.0 $\ifb$) and the dataset used in the previous analysis (2.7 $\ifb$~\cite{y4140}) is shown 
in Figure~\ref{oldvnew}.

\begin{figure}
\begin{minipage}{0.48 \textwidth}
  \begin{center}
\includegraphics[width=7.5cm]{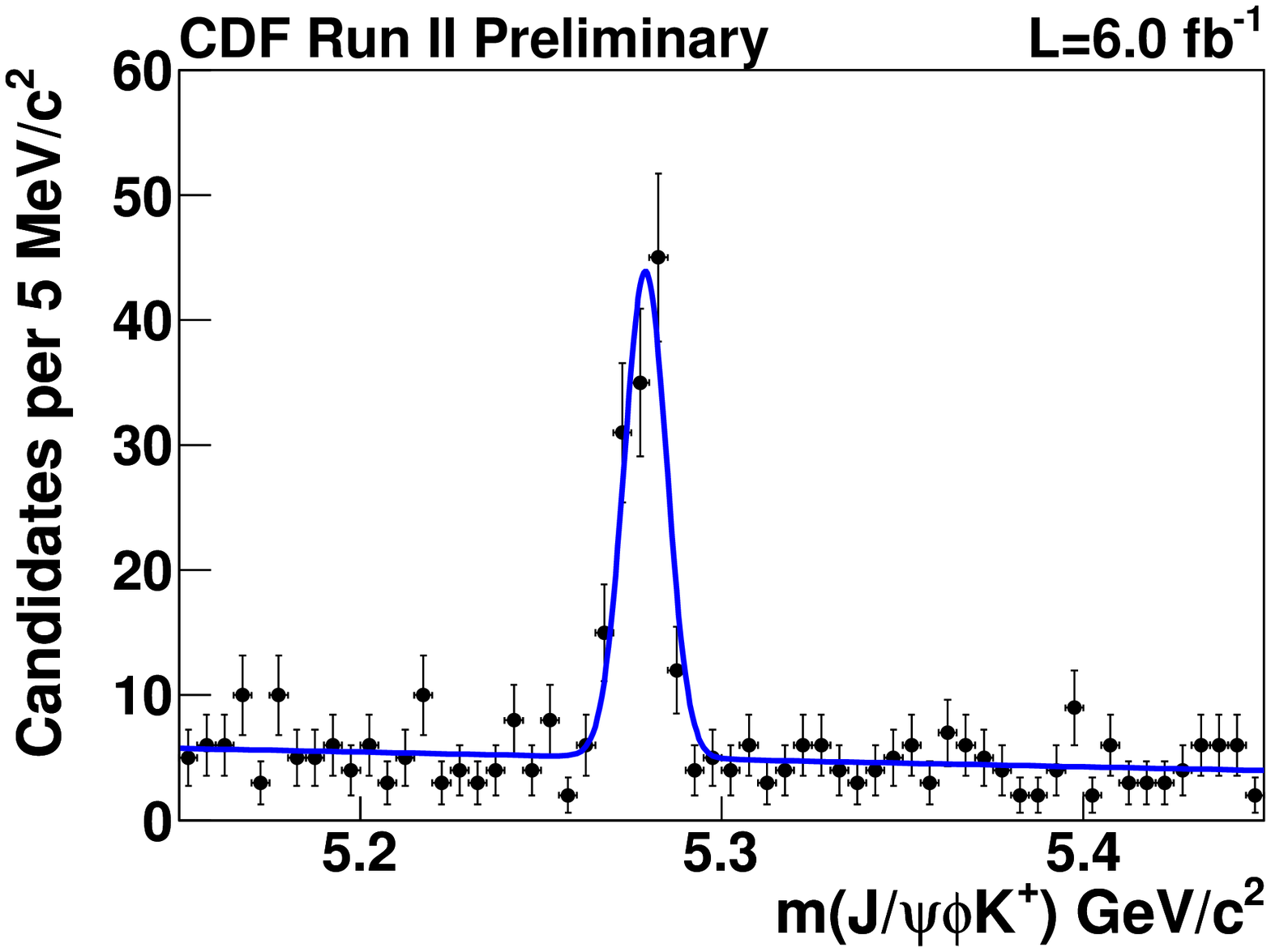}
\caption{\label{myfig1} 
The mass distribution of $J/\psi\phi K^+$;
the solid line is a fit to the data with a Gaussian signal 
function and linear background function.
}
  \end{center}
\end{minipage}
\hspace{0.5cm}
\begin{minipage}{0.48 \textwidth}
  \begin{center}
\includegraphics[width=7.5cm]{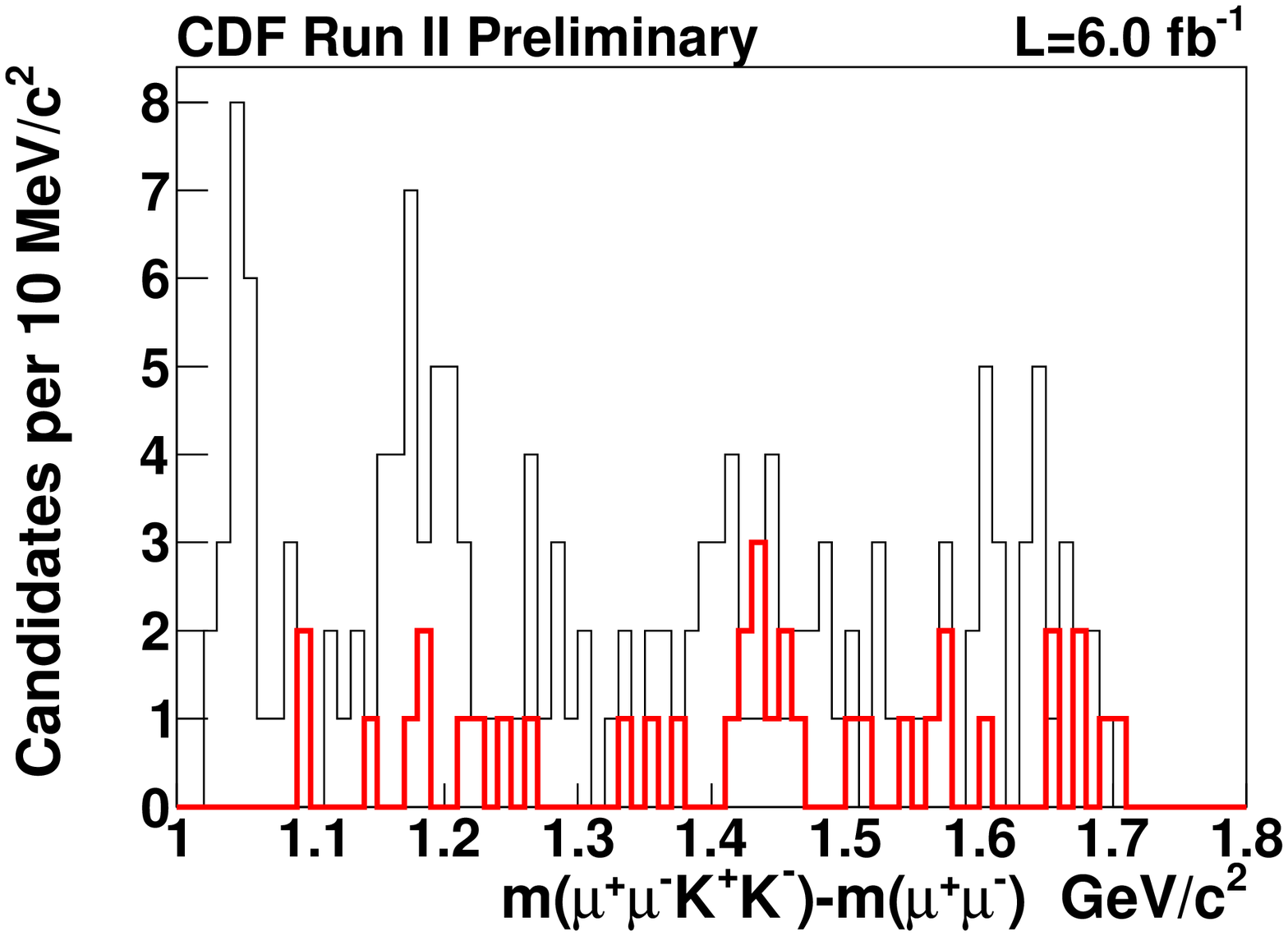}
\caption{\label{myfig2} 
 The mass difference, $\Delta M$, between $\mu^+\mu^-K^+K^-$ and $\mu^+\mu^-$,  
 in the $B^+$ mass window is shown as the black histogram.
The red histogram is the $\Delta M$ distribution 
from the data in the B sideband.
}
  \end{center}
\end{minipage}
\end{figure}

\begin{figure}
\begin{minipage}{0.48 \textwidth}
  \begin{center}
\includegraphics[width=7.5cm]{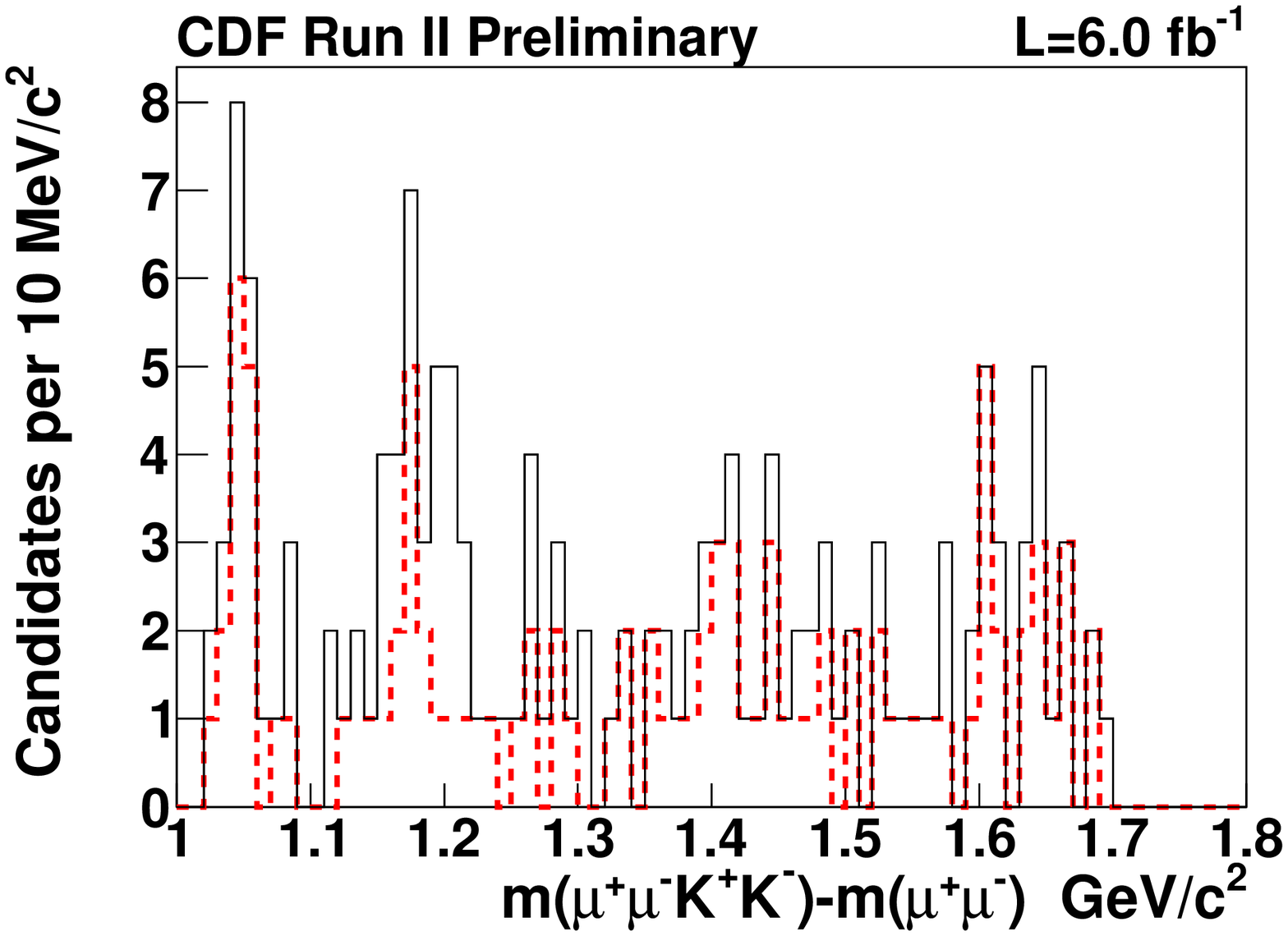}
\caption{\label{oldvnew} 
The $\Delta M$ distribution in the B mass window for the data used in the 
current analysis (6.0 $\ifb$) is 
shown as the black histogram, and the same distribution for the data in the 
previous analysis(2.7 $\ifb$~\cite{y4140})is shown as the red dashed histogram.
}
  \end{center}
\end{minipage}
\hspace{0.5cm}
\begin{minipage}{0.48 \textwidth}
  \begin{center}
\includegraphics[width=7.5cm]{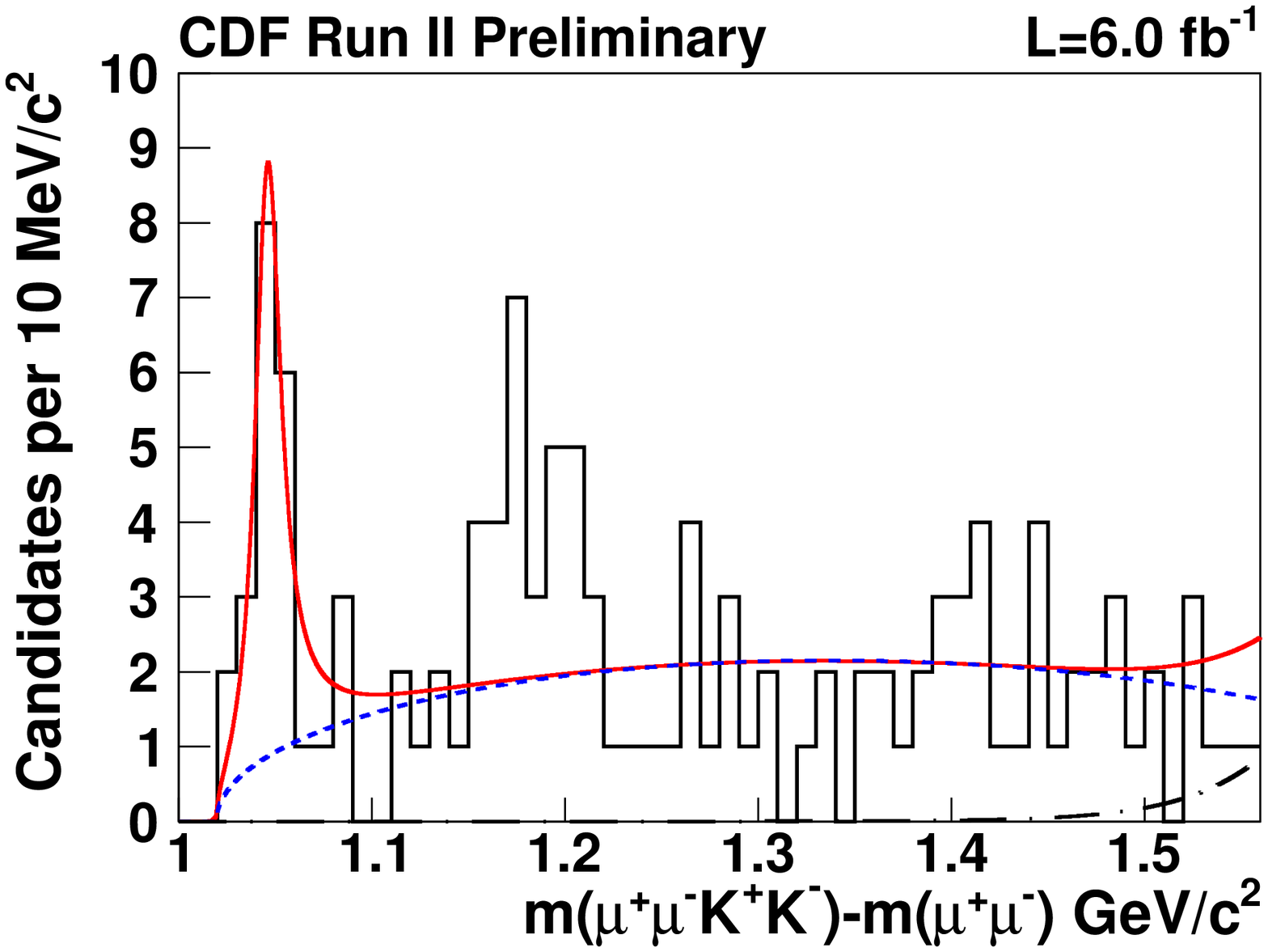}
\caption{\label{dmfit} 
The mass difference, $\Delta M$, between $\mu^+\mu^-K^+K^-$ and $\mu^+\mu^-$, 
in the $B^+$ mass window is shown as a solid black  histogram for the data. 
The dotted curve is the predicted three-body phase space background 
contribution, the dash-dotted curve is 
the predicted $B_s$ contamination (fixed to 3.3), and the solid red curve is the total unbinned fit
where the signal PDF is an S-wave Breit-Wigner convoluted with the resolution (1.7~\MeVcc~). 
}
  \end{center}
\end{minipage}
\end{figure}

The same model is used to examine the $Y(4140)$ structure as described in reference~\cite{y4140}. 
We model the enhancement by an $S$-wave relativistic BW 
function~\cite{sbw} convoluted with 
a Gaussian resolution function with the r.m.s. fixed to 1.7 \MeVcc~obtained from MC,  
and use three--body phase space~\cite{PDG} to describe the background shape.
Even though we exclude the high mass region to avoid the $B_s$ contamination, 
there is still a small contribution in the region of interest. 
We obtained the $\Delta M$ shape from $B_s$ contamination 
and fix the $\Delta M$ shape obtained from $B_s$ MC simulation, 
and the yield to 3.3, scaled from the $B_s\rightarrow J/\psi\phi$ yield in data.
An unbinned likelihood fit to the $\Delta M$ distribution,   
as shown in Fig.~\ref{dmfit}, returns a yield of $\yieldone$ events,   
a $\Delta M$ of $\massdifffitone$, and a width of $\widthfitone$.

\begin{figure}
\begin{minipage}{0.48 \textwidth}
  \begin{center}
\includegraphics[width=7.5cm]{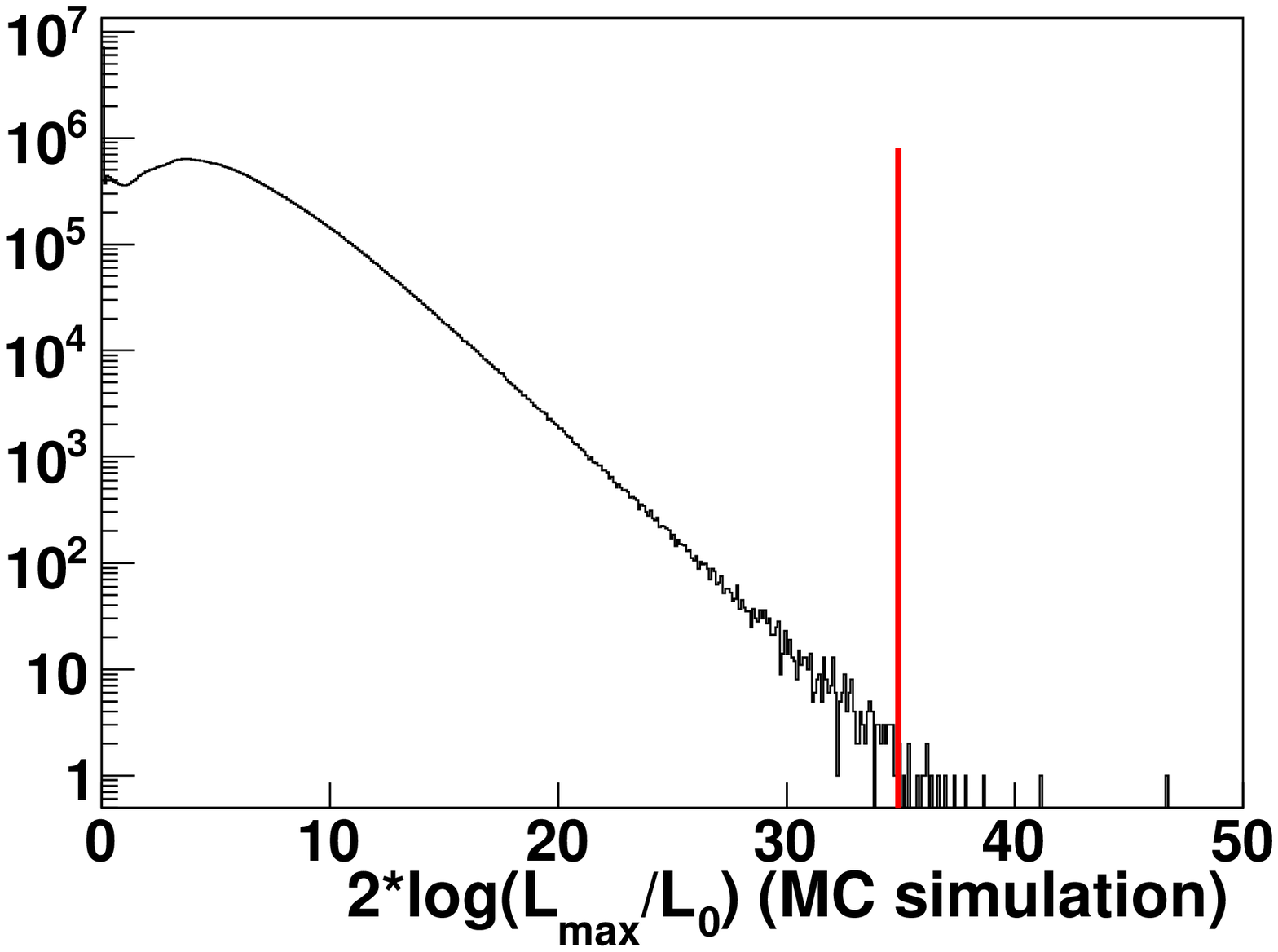}
\caption{\label{logLcount} 
Distribution of $-2{\mathrm{ln}}(\mathcal{L}_0/\mathcal{L}_{{max}})$ for 84 million simulation trials.
The $p$-value obtained from counting is $2.3\times 10^{-7}$, corresponding to a significance of  5.0$\sigma$.
}
  \end{center}
\end{minipage}
\hspace{0.5cm}
\begin{minipage}{0.48 \textwidth}
  \begin{center}
\includegraphics[width=7.5cm]{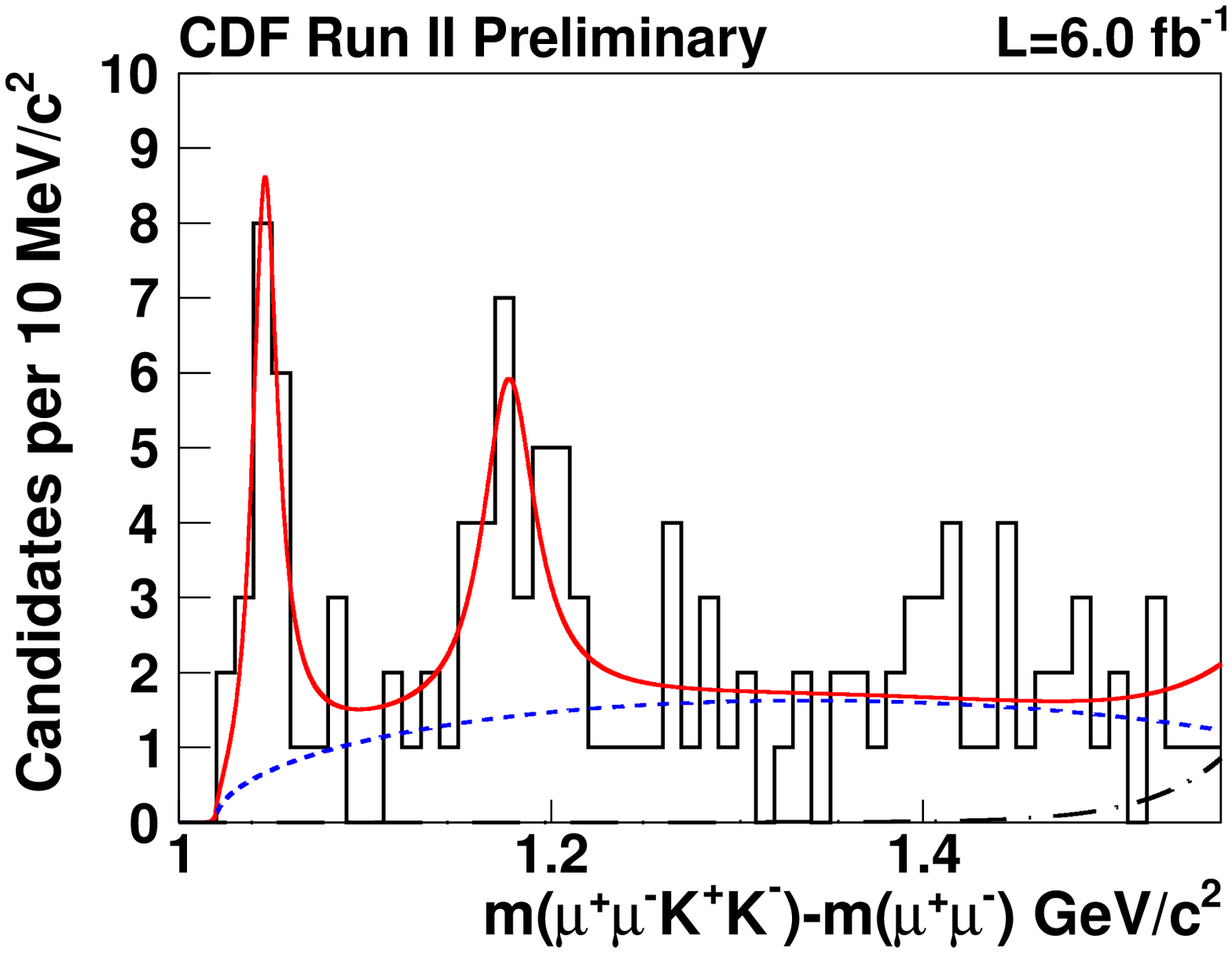}
\caption{\label{twopeaks} 
  The mass difference, $\Delta M$, between $\mu^+\mu^-K^+K^-$ and $\mu^+\mu^-$,  
 in the $B^+$ mass window. 
The dotted curve is the background contribution, the dash-dotted curve is 
the $B_s$ contamination,  and the red solid curve is the total unbinned fit 
assuming two structures. 
}
  \end{center}
\end{minipage}
\end{figure}

We use the same simulation process as in Reference~\cite{y4140}, based 
on a pure three--body phase space background shape to determine the significance 
of the $Y(4140)$ structure. We performed a total of 84 million 
simulations and found 19 trials with a $\sqrt{-2{\mathrm{ln}}(\mathcal{L}_0/\mathcal{L}_{{max}})}$ value 
greater than or equal to the value obtained in the data (5.9), as shown in Fig.~\ref{logLcount}, where $\mathcal{L}_0$ and $\mathcal{L}_{{max}}$ are the likelihood 
values for the null  hypothesis fit and signal hypothesis fit.
The resulting $p$-value  
is $2.3\times 10^{-7}$, corresponding to a significance of greater than 5.0$\sigma$.

The mass of this structure, including systematic uncertainty, 
is measured as $\massone$ after including the world-average $J/\psi$ mass. 
The relative efficiency between~$B^+\rightarrow Y(4140) K^+, Y(4140)\rightarrow J/\psi\phi$
and $B^+\rightarrow J/\psi\phi K^+$ is 1.1 assuming $Y(4140)$ as an S-wave structure 
and $B^+$ phase space decays. Thus the relative branching fraction between 
$B^+\rightarrow Y(4140) K^+$, $Y(4140)\rightarrow J/\psi\phi$
and $B^+\rightarrow J/\psi\phi K^+$ including systematics is $\rebfone$.

An further excess above the three-body phase space background shape appears at approximately 1.18 \GeVcc~ in Fig.~\ref{myfig1} (b). 
Since the significance of $Y(4140)$ is greater than 5$\sigma$, we fit to the data 
assuming two structures at $\Delta M$ of 1.05 and 1.18 \GeVcc~ as shown in 
Fig.~\ref{twopeaks}. 
The fit to the data with the same requirements as in the previous paper~\cite{y4140} 
returns a yield of $20\pm5$ events,   
a $\Delta M$ of $1046.7^{+2.8}_{-2.9}$ \MeVcc~, and a width 
of $15.0^{+8.5}_{-5.6}$ \MeVcc~ for the $Y(4140)$, which are consistent with 
the values returned from a $Y(4140)$-only signal fit.
The fit returns a yield of $22\pm8$ events,   
a $\Delta M$ of $1177.7^{+8.4}_{-6.7}$ \MeVcc~, and a width 
of $32.3^{+21.9}_{-15.3}$ \MeVcc~ for the structure around $\Delta M$ of 1.18 \GeVcc~.
The significance of the second structure, determined by a similar 
simulation is 3.1$\sigma$.

In summary, the growing $B^+ \rightarrow J/\psi \phi K^+$ sample at CDF 
enables us to observe the $Y(4140)$ structure~\cite{y4140} 
with a significance greater than 5$\sigma$. Assuming an $S$-wave relativistic BW, 
the mass and width of this structure, 
including systematic uncertainties, are measured to be $\massone$ 
and $\widthone$, respectively. The relative branching fraction between 
$B^+\rightarrow Y(4140) K^+, Y(4140)\rightarrow J/\psi\phi$
and $B^+\rightarrow J/\psi\phi K^+$ including systematics is $\rebfone$.
We also find evidence at 3.1$\sigma$ level for a second structure with a 
mass of $\masstwo$, a width of $\widthtwo$ and a yield of $\yieldtwo$.


\begin{thebibliography}{99}


\bibitem{y4140}
T. ~Aaltonen  { \em et al.}~(CDF Collaboration), Phys.\ Rev.\ Lett. {\bf 102}, 242002 (2009) .





\bibitem{y4140theory}
X.~Liu and S.~Zhu, Phys.\ Rev.\ D\ {\bf 80}, 017502~(2009);
N.~ Mahajan, Phys.\ Lett.\ B\ {\bf 679}, 228~(2009); 
Z.~Wang, Eur.\ Phys.\ J.\ C\ {\bf 63}, 115~(2009);
T.~ Branz {\em et al.}, Phys.\ Rev.\ D {\bf 80}, 054019~(2009);
R.~ Albuquerque {\em et al.}, Phys.\ Lett.\ B\ {\bf 678}, 186~(2009);
X.~Liu, Phys.\ Lett.\ B\ {\bf 680}, 137~(2009);
G~ Ding, Eur.\ Phys.\ J.\ C\ {\bf 64}, 297~(2009);
J.~ Zhang and M.~ Huang, Phys.\ Rev.\ D\ {\bf 80}, 056004~(2009);
E.~ Beveren and  G.~ Rupp, arXiv:0906.2278 [hep-ph];
F.~ Stancu, J.\ Phys.\ G\ {\bf 37}, 075017~(2010);
T.~ Branz {\em et al.}, arXiv:1001.3959 [hep-ph];
K.~ Yamada, arXiv:1002.0410 [hep-ph].




\bibitem{cdfii}
D.~Acosta  {\em et al.} (CDF Collaboration), Phys.\ Rev.\ D {\bf 71},  032001 (2005);
A.~Abulencia {\em et al.} (CDF Collaboration), Phys.\ Rev.\ Lett. {\bf 97}, 242003 (2006).




\bibitem{cdfbc}
A.~Abulencia {\em et al.} (CDF Collaboration), Phys.\ Rev.\ Lett. {\bf 96}, 082002 (2006);
T.~Aaltonen {\em et al.} (CDF Collaboration), Phys.\ Rev.\ Lett. {\bf 100} 182002 (2008).



\bibitem{sbw}
$\frac{dN}{dm} \propto \frac{m \Gamma(m)}{(m^2-m_0^2)^2+m_0^2\Gamma^2(m) }$, where
$\Gamma(m)=\Gamma_0\frac{q}{q_0}\frac{m_0}{m}$, and the 0 subscript indicates the 
value at the peak mass.



\bibitem{PDG}
C.~Amsler {\it et al.}  ~(Particle Data Group), Phys.\ Lett.\ B {\bf 667}, 1 (2008).




\end{thebibliography}
\end{document}